\documentclass[sigconf]{acmart}
\usepackage{graphicx}
\usepackage{subfig}
\AtBeginDocument{%
  }

\setcopyright{acmlicensed}
\copyrightyear{2025}
\acmYear{2025}
\setcopyright{acmlicensed}\acmConference[UbiComp Companion
'25]{Companion of the 2025 ACM International Joint Conference on Pervasive
and Ubiquitous Computing}{October 12--16, 2025}{Espoo, Finland}
\acmBooktitle{Companion of the 2025 ACM International Joint Conference on
Pervasive and Ubiquitous Computing (UbiComp Companion '25), October
12--16, 2025, Espoo, Finland}
\acmDOI{10.1145/3714394.3756193}
\acmISBN{979-8-4007-1477-1/2025/10}

\acmSubmissionID{6884}

\begin{document}

\title[Robust In-the-Wild Exercise Recognition from a Single Wearable]{Robust In-the-Wild Exercise Recognition from a Single Wearable: Data-Side Fusion, Sensor Rotation, and Feature Engineering}

\author{Hoang Khang Phan}
\email{khang.phan2411@hcmut.edu.vn}
\orcid{0009-0007-1578-0977}
\authornote{The authors contribute equally in this study.}
\affiliation{%
\department{Department of Biomedical Engineering}
  \institution{Ho Chi Minh City University of Technology (HCMUT), VNU-HCM}
  \city{Ho Chi Minh City}
  \country{Vietnam}
}

\author{Khang Le}
\orcid{0009-0006-7162-6045}
\email{khang.le2710@hcmut.edu.vn}
\authornotemark[1]
\affiliation {%
\department{Faculty of Computer Science and Engineering}
  \institution{Ho Chi Minh City University of Technology (HCMUT), VNU-HCM}
  \city{Ho Chi Minh City} 
  \country{Vietnam}
}

\author{Tu Nhat Khang Nguyen}

\email{khang.nguyentunhat@gmail.com}
\orcid{0009-0007-9957-1178}
\affiliation{%
\department{Faculty of Computer Science and Engineering}
  \institution{Ho Chi Minh City University of Technology (HCMUT), VNU-HCM}
  \city{Ho Chi Minh City}
  \country{Vietnam}
}

\author{Anh Van Dao}
\orcid{0009-0008-2103-4599}
\email{anh.dao1606@hcmut.edu.vn}
\affiliation{%
\department{Department of Applied Mathematics}
  \institution{Ho Chi Minh City University of Technology (HCMUT), VNU-HCM}
  \city{Ho Chi Minh City}
  \country{Vietnam}
}

\author{Nhat Tan Le}

\email{lenhattan@hcmut.edu.vn}
\orcid{0000-0002-2738-6607}
\authornote{Corresponding author.}
\affiliation{%
\department{Department of Biomedical Engineering}
  \institution{Ho Chi Minh City University of Technology (HCMUT), VNU-HCM}
  \city{Ho Chi Minh City}
  \country{Vietnam}
}


\begin{abstract}
Monitoring physical exercises is vital for health promotion, with automated systems becoming standard in personal health surveillance. However, sensor placement variability and unconstrained movements limit their effectiveness. This study proposes the team "3KA"'s one-sensor workout activity recognition method using feature extraction and data augmentation in $2^{nd}$WEAR Dataset Challenge. From raw acceleration, angle and signal magnitude vector features were derived, followed by extraction of statistical, fractal/spectral, and higher-order differential features. A fused dataset combining left/right limb data was created, and augmented via sensor rotation and axis inversion. We utilized a soft voting model combining Hist Gradient Boosting with balanced weights and Extreme Gradient Boosting without. Under group 5-fold evaluation, the model achieved 58.83\% macro F1 overall (61.72\% arm, 55.95\% leg). ANOVA F-score showed fractal/spectral features were most important for arm-based recognition but least for leg-based.
The  code to reproduce the experiments is publicly available via: \href{https://github.com/Khanghcmut/WEAR_3KA}{https://github.com/Khanghcmut/WEAR\_3KA}

\end{abstract}




\maketitle
\section{Introduction}
Physical training has been essential in human development and long-term survival throughout history by contributing to health maintenance and enhancing physiological endurance. In the modern era, the automated tracking of physical activity is emerging as a new standard in personal health monitoring. Initially, smartphones were explored for this purpose \cite{weiss2016smartwatch,gyllensten2011identifying}. Shoaib et al. \cite{shoaib2013towards} successfully used smartphone sensors to classify six different walking-related activities in a constrained environment. By using two simple statistical features - mean and standard deviation, this research reached over 95\% TPR in activity recognition. Similarly, studies by Ciliberto et al. \cite{ciliberto2017high,wang2021three} utilized 4 smartphones on various body parts to distinguish between physical activities like walking and cycling, as well as different modes of transport. Utilizing Ciliberto study's dataset, Zhu et al. \cite{zhu2020densenetx} applied DenseNetX deep learning architecture, which reached up to 84\% F1 score in type of locomotion detection.

However, these studies also highlight the inherent limitations of smartphones for detailed analysis. While these devices are effective in detecting general movements, inconsistencies in their placement often limit their ability to accurately capture the finer details of complex activities. This limitation has led to the increased adoption of smartwatches for complex human activity recognition (HAR). Worn securely at a precise location like the wrist, smartwatches minimize signal noise caused by extraneous movement and are better suited for capturing the fine-grained data required to analyze complex exercises. This advantage was demonstrated by Weiss et al. \cite{weiss2019smartphone}, who compared smartphone and smartwatch data for recognizing 18 activities in laboratory conditions. Their research confirmed that the smartwatch was superior for identifying intricate actions like clapping and catching, reinforcing its suitability for complex activity recognition.

Nonetheless, one of the most significant yet underdeveloped issues in wearable computing studies is the inherent variability in sensor placement. In real-world, unconstrained environments, the precise orientation and position of wearable devices are rarely guaranteed. Participants may inadvertently wear the device improperly - for instance, on the wrong intended limb side, upside-down, or shifted from its intended location, or simply adjust it for personal comfort. These deviations from the ideal placement create a multitude of "unseen scenarios" for a machine learning model. When a model encounters data from these novel orientations, which differ significantly from the carefully collected training data, its predictive performance can degrade drastically. This challenge severely limits the practical deployment and reliability of activity recognition systems, creating a critical need for methods that are resilient to such real-world inconsistencies.

To address this fundamental problem, this study is designed to develop and validate a robust methodology for exercise recognition that is less susceptible to the challenges of sensor misplacement. The primary aims of our research are therefore twofold:
\begin{enumerate}
    \item Extract a robust set of features from accelerometer data for exercise activity recognition.
    \item Leverage the extracted features for cross-limb side exercise activity recognition.
\end{enumerate}

\section{Methodology}

\begin{figure*}[!ht]
    \centering
    \includegraphics[width=0.77\linewidth]{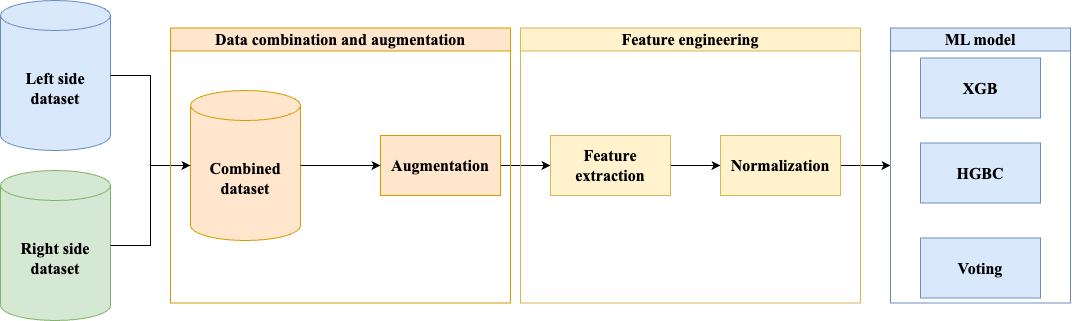}
    \caption{The flowchart of the proposed pipeline. For each limb, data from the right and left sides were first combined. Data augmentation was then applied to the merged data, followed by feature engineering. Finally, a single model was trained per limb, making the pipeline practical for real-world use.}
    \label{pipeline}
\end{figure*}
In this study, we utilized accelerometer data in each limb for physical workout recognition (see Figure \ref{pipeline}). Initially, we fused the data from left and right sides of each limb to create a limb-specific dataset. After that, we augmented the data based on possible real-world data scenarios. Finally, a total of 450 features were extracted for exercise activity recognition.
\subsection{Dataset}
The dataset utilized for the WEAR challenge is a comprehensive collection of human activity data, originating from the extensive WEAR dataset \cite{bock2024wear}. This rich dataset captures the nuances of 18 distinct exercise activities, thoughtfully categorized into three fundamental groups: jogging, stretching, and strength training. To ensure a diverse and robust dataset, data were collected from 26 participants. The data acquisition process was meticulous, employing a multi-modal approach. High-fidelity accelerometer data was captured using four Bangle.js smartwatches, with one securely attached to each wrist and ankle of the participants. In addition to the inertial sensor data, synchronized video recordings were captured for the entire duration of each exercise session. These videos were recorded using either a GoPro Hero 8 or a Hero 11 action camera, ensuring high-resolution visual context for the physical activities.

For the aims of the WEAR challenge, this dataset was partitioned into distinct training and testing sets to rigorously evaluate the performance of participating models. The training set consisted of the raw, unsegmented accelerometer data from 22 of the participants. This allows teams to develop and refine their models on a substantial body of information. In contrast, the test dataset was strategically designed to assess the real-world generalizability and robustness of the developed models. This test set comprises data from the remaining four participants and was deliberately segmented into one-second windows. Furthermore, this segmented data underwent augmentation, introducing variations and complexities intended to simulate real-life scenarios and challenge the models' ability to adapt to unfamiliar conditions.

\subsection{Data Augmentation Strategy} \label{DC}

A significant challenge in wearable sensor studies is the lack of standardized placement, as the orientation and exact position of a sensor often vary from person to person due to individual preferences and comfort.
During our data analysis, we encountered direct evidence of this challenge. For instance, we observed that participants 2 and 5 wore their devices on the incorrect wrist, resulting in a left-right swapping of the data channels. This type of error can fundamentally alter the interpretation of asymmetrical movements. Furthermore, participants 9 and 18 wore the left sensor in an upside-down orientation. This specific misplacement inverts the sensor's coordinate system, causing, for example, the accelerometer on the x-axis to be flipped in case of participants 2 and 5's wrist accelerometer data, which severely compromises the integrity of the raw accelerometer readings. Such inconsistencies in sensor deployment present a substantial obstacle to developing robust and generalizable activity recognition models. Therefore, in this study, we propose and implement two such techniques, including axis inversion and sensor rotation, to enrich our dataset.

These augmentation strategies are detailed below:
\begin{itemize}
    \item \textbf{Axis inversion}: This technique simulates the common scenario where a user wears a device on the opposite limb from which it was trained (e.g., wearing a device on the left arm when the model expects it on the right). Such a swap often results in a mirror image of the motion signal along a primary axis. To replicate this, we applied a strategic inversion to the data: the x-axis signal was inverted for the arm-worn sensor data, and the y-axis signal was inverted for the leg-worn sensor data.
    \item \textbf{Sensor rotation}: It is also common for a user to wear a device upside-down (e.g., with a watch face oriented incorrectly). Noticing that this would fundamentally alter the sensor's coordinate system, we augmented the data to account for this possibility. The scenario was simulated by applying a 180-degree mathematical rotation to the sensor data around its x-axis, effectively generating data equivalent to an upside-down placement.
\end{itemize}

\subsection{Feature extraction}

\begin{table}[!ht]
    \centering
    \caption{The table of feature type and quantities.}
    \begin{tabular}{cc}
    
    \toprule
    \textbf{Feature type}& \textbf{Quantities} \\
         \hline
         Raw-accelerometer-based features& 135  \\

         Signal-Magnitude-Vector-based features & 180\\

         Angle-based & 135 \\
         \bottomrule
    \end{tabular}
    
    \label{quantities} 
\end{table} 
\subsubsection{Raw-accelerometer-based features}
Accelerometer data is a cornerstone of Human Activity Recognition (HAR) as it captures the rich, nuanced details of physical motion. The utility of this data is twofold. Firstly, the dynamic component of the accelerometer signal represents the movement behavior itself. Analyzing features related to this component, such as the signal's magnitude and rate of change, provides profound insight into the way each exercise is performed, enabling the differentiation of complex limb movements. Secondly, the raw accelerometer data reveals limb orientation through its alignment with the Earth's gravitational field, a constant downward force of 1g. This static component acts as a stable reference vector. For example, when a participant is lying down to perform a sit-up, the y-axis of an accelerometer placed on the ankle would align with the downward gravitational vector. In contrast, during standing-based activities, the x-axis of a sensor on the hand might align with gravitational orientation, thus providing crucial contextual information about the participant's posture.

To comprehensively capture these dynamic and static characteristics from the raw signal, a robust feature engineering process was employed. In this study, we extracted a total of 45 features for each axis of the raw accelerometer data. These features were drawn from several domains to create a rich representation of the activities. The extracted features consist of:
\begin{itemize}
    \item Statistical and Temporal Features: A comprehensive set of 27 features was derived using the TSFEL library and other common metrics. This includes fundamental statistical descriptors such as Mean, Median, Mode, Max, Min, Standard Deviation, and Variance, which summarize the signal's distribution. It also includes features that describe the signal's energy and structure, such as Interquartile range, Root Mean Square (RMS), Average Power, Absolute Energy, Peak-to-Peak distance, mean crossing rate, AUC, and Entropy. The full set of temporal features from the TSFEL library \cite{barandas2020tsfel} was also included to capture characteristics of the signal over time.

\item Fractal and Spectral Features: To quantify the complexity and periodicity of the movements, we included 4 features: two fractal dimension features (Petrosian and Katz) and two spectral features (Top Dominant Frequency and Second Dominant Frequency). These features help distinguish between simple, repetitive motions and more complex, irregular activities.

\item Higher-Order Differential Features: To analyze the rate of change in motion (jerk and jounce), the raw signal was differentiated twice. From the second ($d^2s$) and third ($d^3s$) order differentials (see Equation \eqref{df}), 14 features consist of Mean, Median, Standard Deviation, Mean Absolute, Median Absolute, Standard Deviation Absolute, and Katz fractal dimension were calculated. The differentiation process is defined by the equation:
\begin{equation}
d^{n}s=\Delta(d^{n-1} s)
\label{df}
\end{equation}
where n is the rank of the differential.

\end{itemize}

\subsubsection{Signal-Magnitude-Vector-based features}
The Signal Magnitude Vector (SMV) is a crucial and widely used feature in signal processing, particularly for analyzing data from multi-axis sensors such as accelerometers and gyroscopes. It provides a comprehensive, cross-axis measure of the overall magnitude or intensity of a signal at any given point in time.

In applications such as Human Activity Recognition (HAR), a primary challenge is the variability in sensor orientation. For instance, a sensor positioned on a user's wrist or carried in a pocket may exhibit a different orientation compared to when the user raises their hand. Consequently, the raw accelerometer readings along the x, y, and z axes for an identical physical movement can vary significantly depending on the device's orientation. This orientation dependency introduces variability that may confuse machine learning models, leading to poor performance and a lack of generalizability.

The SMV solves this problem by creating a single, orientation-independent data stream. It achieves this by calculating the Euclidean norm of the vector formed by the individual axes. For a three-axis accelerometer with readings (x,y,z), the SMV at each time point is calculated as \eqref{SMV}:

\begin{equation}
    SMV=\sqrt{x^2+y^2+z^2}
    \label{SMV}
\end{equation}

Additionally, two-axis SMV features were computed to enhance the comprehensiveness of the feature representation (see Equation \eqref{SMV2}).
\begin{equation}
    SMV_{u,v}=\sqrt{u^2+v^2}
    \label{SMV2}
\end{equation}
 where $u$ and $v$ is 2 of $x,\;y,$ or $ z$ axis.
 
In this study, to reduce computational complexity, we computed the squared form of the SMV. Subsequently, 45 time-series-based features were extracted for each modality of this feature set.
\subsubsection{Angle-based features}
While Signal Magnitude Vector (SMV) addresses the overall intensity irrespective of orientation, it does not fully capture the nuanced positional changes and specific limb orientations critical for comprehensive activity recognition. To address this limitation, we incorporate angle-based features, offering a powerful complement by providing contextual information about the relative orientation of the movement. This makes angle-based features highly valuable for distinguishing between activities that involve similar overall motion intensity but differ significantly in their posture or limb angles.

In this study, we leveraged the relationship between pairs of accelerometer axes to derive angle-based features. Specifically, for each pair of axes $(u, v)$ (where $u$ and $v$ are two of the $x, y, z$ axes), the angle $\theta_{uv}$ was calculated using the formula \eqref{ang}:

\begin{equation}
\theta_{uv} =  \arctan2(u, v)
\label{ang}
\end{equation}

This resulted in three angular time series: ($\theta_{xy}$, $\theta_{xz}$, and $\theta_{yz}$). From each of these, we extracted 45 time-series features to capture the dynamic patterns of relative axes. 

\subsection{Feature Normalization}

In machine learning applications, feature normalization plays a pivotal role in model performance and accurate recognition. The selection of a suitable normalizer is critical, as it not only standardizes the data into a uniform range but can also reduce the influence of outliers. For this research, we utilized the \texttt{QuantileTransformer} from the scikit-learn library, which transforms features to follow a uniform or normal distribution, making it robust to extreme values \cite{pedregosa2011scikit}.

\subsection{Classification Model}
We employed the following machine learning models for classification:

\begin{itemize}
    \item \textbf{Histogram-based Gradient Boosting Classifier (HGBC) \cite{hgbc}:} An efficient implementation of gradient boosting that discretizes continuous features into histograms to speed up training. It is particularly well-suited for large datasets and supports native handling of missing values. In this study, HGBC was trained with balanced class weights.
    
    \item \textbf{Extreme Gradient Boosting (XGB) \cite{XGB}:} A scalable and regularized version of gradient boosting that uses advanced optimization techniques such as parallelization, tree pruning, and sparsity-aware algorithms to achieve high performance and prevent overfitting. In this study, XGB was not trained with balanced class weights.

    \item \textbf{Soft Voting Ensemble:} Classification strategy relying on both XGB and HGBC, each trained using the configurations described above. The final prediction was computed as the average of the class probabilities output by the two models.
\end{itemize}

\subsection{Evaluation settings}

To evaluate generalization to new individuals, we used group 5-fold cross-validation based on participant ID. The 22 training participants were split into five non-overlapping groups; in each fold, the model was trained on four groups and validated on the remaining one. This subject-wise split prevents data leakage and ensures the model is tested only on unseen individuals, providing a realistic estimate of real-world performance by avoiding overfitting to individual movement patterns.

To evaluate model performance, the macro F1-score was selected as the primary performance metric. This metric is particularly well-suited for multi-class problems, especially those with potential class imbalances, as it calculates the F1 score for each class independently before taking their unweighted average. By giving equal importance to each category, regardless of its frequency in the dataset, the macro F1 score provides a comprehensive and unbiased assessment of the model's effectiveness across all classes.

\section{Results}
The experiments were conducted on a 14-inch MacBook Pro equipped with an Apple M1 Pro chip with 8 CPU cores and 16 GPU cores. The complete pipeline required approximately 24 hours of execution. 

\subsection{Classification performance}
Table \ref{classify_res} presents the classification performance of the three evaluated models in recognizing exercise activities, measured by the macro F1-score. A notable observation is that the models consistently achieved better results using data from the arm-worn accelerometer, outperforming the leg-based sensor by a margin of approximately 6 to 8 percentage points in the macro F1-score. These results suggest that, for the exercises studied, sensor placement on the arm provides more discriminative motion patterns for classification.
\begin{table}[!ht]
\caption{The macro F1 score of classification for each limb (in percent)}
    \centering
    \begin{tabular}{c|ccc}
    \toprule
         &\textbf{XGB}&\textbf{HGBC}&\textbf{Voting}  \\
         \hline
        Arm &60.25\textpm2.52 & 58.73\textpm2.92 &\textbf{61.72\textpm3.05 } \\
        Leg& 54.40\textpm 3.49& 52.99 \textpm3.96   & \textbf{55.95\textpm3.67}  \\
        \hline
        Average&  57.22\textpm 3.00 & 55.86\textpm3.44  & \textbf{58.84\textpm 3.36} \\
        \bottomrule
    \end{tabular}
    \label{classify_res}
\end{table}

\begin{figure*}[!ht]
\subfloat[Arm]{\label{cf_arm}
    \centering
    \includegraphics[width=0.456\linewidth]{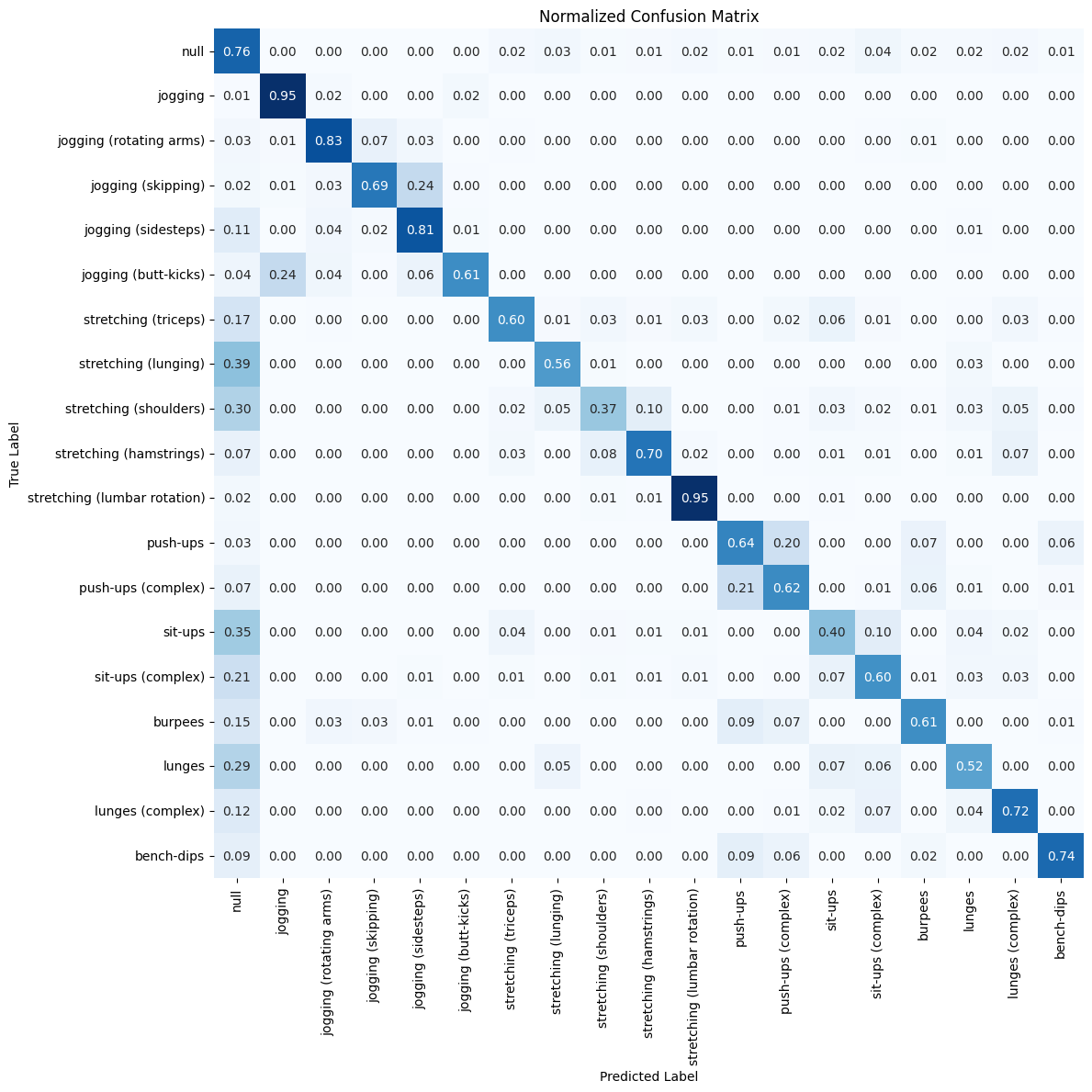}
    }
\subfloat[Leg]{ \label{cf_leg}
    \includegraphics[width=0.456\linewidth]{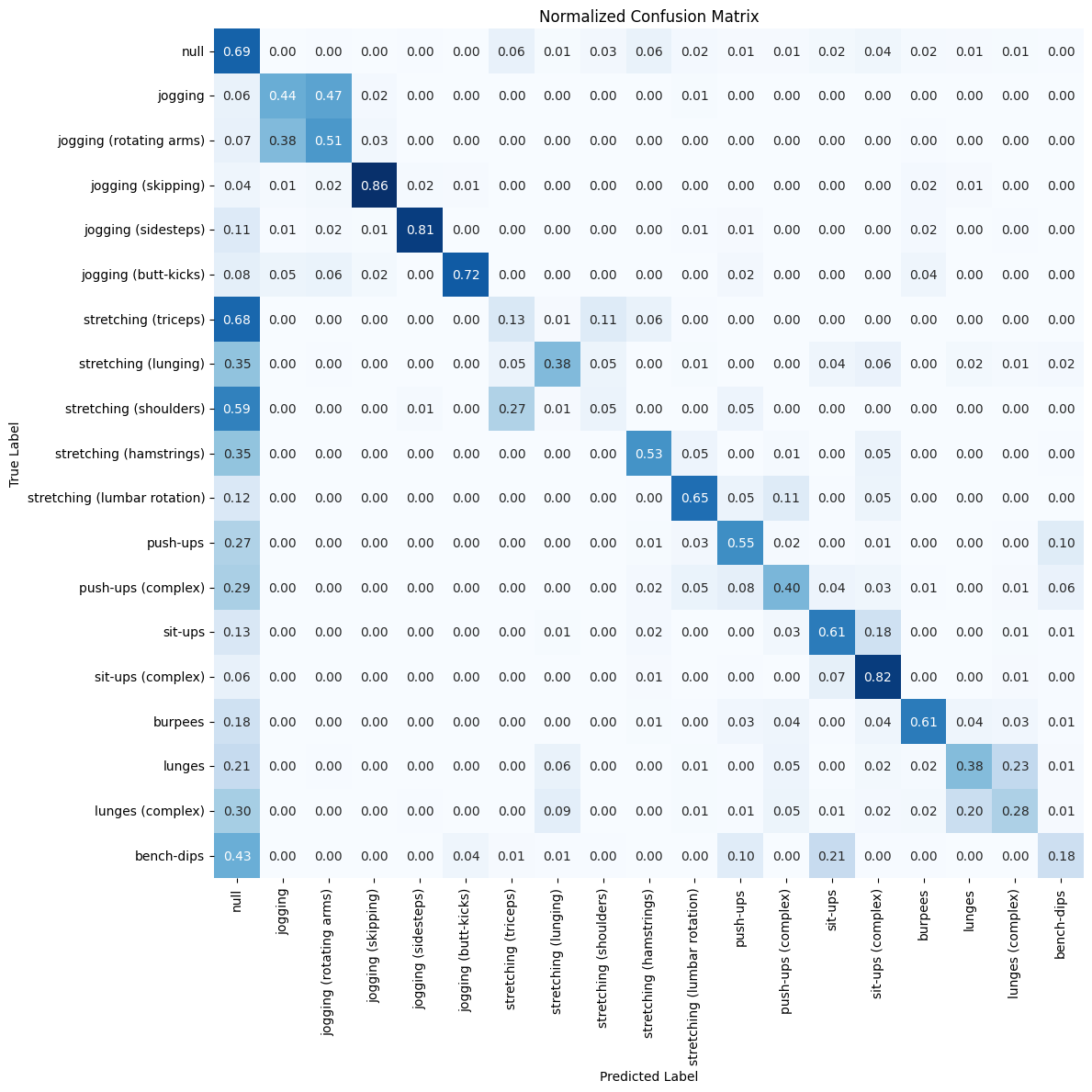} 
    }
   \caption{The confusion matrix of activity recognition by (a) arm and (b) leg.}
    \label{cf}
\end{figure*}
\subsection{Feature analysis}

Among the methods evaluated, the Voting Classifier achieved the highest performance, surpassing the individual models by up to 3-4\% in F1-score across the tested scenarios. This superior performance can be attributed to the synergistic nature of the ensemble, which combines the strong predictive power of the XGB model with the robustness of the HGBC, particularly in handling class imbalances.

\begin{figure}[!ht]
    \centering
    \includegraphics[width=0.75\linewidth]{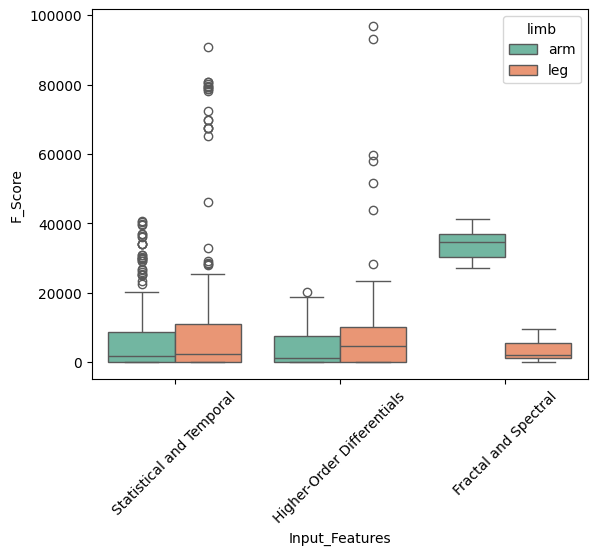}
    \caption{The box plot of ANOVA F score by group of feature}
    \label{F_score}
\end{figure}

This effective fusion allows the model to better navigate one of the key challenges in this dataset: the high similarity between certain activities. Specifically, it improves the differentiation between the 'Null' class (representing the majority of standing or sitting still activity) and various stretching activities. Given that many stretching activities involve maintaining static postures with only subtle limb movements, they can be easily confused with inactivity. The ensemble approach appears better equipped to resolve this ambiguity, leading to a more accurate and reliable classification overall.

Figure \ref{F_score} illustrates the F-scores of the three feature groups for both arm- and leg-based sensors. A key finding is the significance of the "Fractal and Spectral" features. For arm-based recognition, this feature group holds the highest importance, while for leg-based recognition, it is the least informative feature group among all groups considered. Specifically, the F-score for the arm's fractal and spectral features is approximately three times higher than that of the leg's. This suggests that arm movements exhibit more distinct features, which are well-captured by these features. In contrast, leg activity appears to be more consistent and can overlap between different exercises, making these specific features less discriminative.

It is also noteworthy that the "Statistical and Temporal" and "Higher-Order Differentials" feature groups yield slightly higher F-scores for leg-based sensor data compared to arm-based data. This observation can be attributed to the inherent consistency of leg movements in many activities. For instance, while a person can swing their arms freely and variably during a run, leg movements in jogging follow a more constrained and repetitive pattern to maintain balance and forward motion. As a result, "Statistical and Temporal" and "Higher-Order Differentials" are more effective in capturing the nuances of the movement compared to more varied and overlaps from arm-locate data.

\section{Discussion}

This research introduced a cross-side limb data fusion strategy for exercise recognition, yielding promising results with our voting method achieving a macro F1-score of 59\% \textpm 3\%. Moreover, this data fusion strategy offers a potential solution to the challenge of inconsistent sensor placement in unconstrained, real-world environments using wearable devices. However, a key limitation is the model's confusion between normal and complex activities. We hypothesize this occurs because complex activities often begin with movements identical to a normal activity. Consequently, a one-second analysis window may be too short to capture the full, distinguishing sequence, causing segments to be misidentified.

Additionally, when using only leg accelerometer data, the model often misclassifies activities with minimal leg movement (e.g., stretching or bench dips) as "Null" (refer to Figure \ref{cf_leg}). This issue likely stems from the similarity in leg sensor readings during such static activities and requires further investigation.

\begin{figure}[!ht]
\subfloat[Subject 15]{\label{F_arm}
    \centering
    \includegraphics[width=0.45\linewidth]{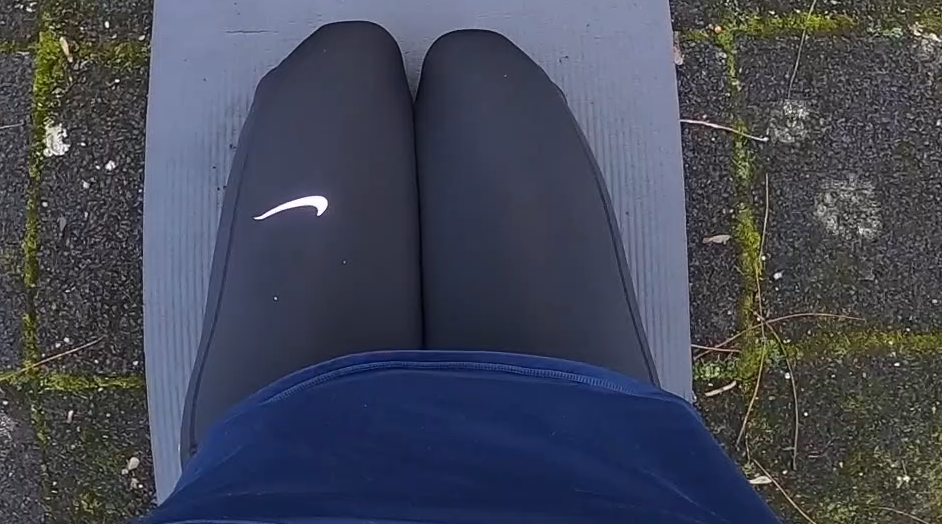}
    }
\subfloat[Subject 18]{ \label{F_leg}
    \includegraphics[width=0.45\linewidth]{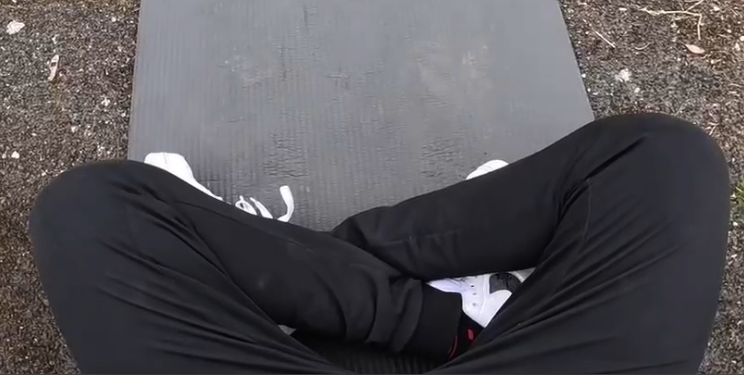} 
    }
   \caption{The camera footage of (a) subject 15 and (b) subject 18 during their abnormal position triceps stretching activity session.}
    \label{fig:activity}
\end{figure}

Furthermore, in the wild, the execution of any given action is not uniform but is instead subject to a wide array of variations. This inherent unpredictability of real-life data stems from the multitude of ways an individual can perform a single activity.  A clear example can be observed in the performance of triceps stretching. Participants 15 and 18, for instance, were recorded performing this activity from a seated position, a modification that can alter the biomechanics of the stretch (see Figure \ref{fig:activity}). In contrast, participants 20, 9, and 7 were observed to be walking while simultaneously executing the same triceps stretch, introducing a dynamic element to the exercise. This demonstrates that even within a controlled study, the natural diversity of human movement persists, highlighting a key challenge in standardizing and analyzing activity recognition data.

Finally, to significantly enhance recognition accuracy, a multi-modal approach is recommended. A primary challenge remains the inconsistent orientation of sensors. This could be resolved by integrating gyroscope data to standardize accelerometer readings to a fixed global coordinate system. For instance, if the gyroscope detects that the sensor's local y-axis is aligned with the global z-axis, the y-axis data can be remapped to the z-axis, thereby correcting for sensor rotation during an activity.

\section{Conclusion}
This study presented team "3KA" solution for exercise activity recognition , which utilizes a strategy of combining accelerometer data from contralateral limbs, at 2025 HASCA WEAR Challenge. This approach achieved a recognition accuracy as high as 61.72\%. Furthermore, our data analysis identified key challenges inherent to real-world data collection: significant variations in exercise execution and inconsistent sensor placement among the participants. These insights into the nature of "in-the-wild" data are valuable, suggesting a clear path for future research. Specifically, generative models could be employed to synthesize data that captures the diverse ways an activity might be performed, thereby enhancing model robustness and performance.
\begin{acks}
We acknowledge Ho Chi Minh City University of Technology (HCMUT), VNU-HCM for supporting this study.
\end{acks}

\bibliographystyle{ACM-Reference-Format}
\bibliography{3KA}

\end{document}